\documentclass[cite1,clbiba,12pt,psfig]{article}
\usepackage{epsfig}
\usepackage{graphicx}
\usepackage{overcite}
\usepackage{chapterbib}

\usepackage{fancyhdr}
\usepackage{hyperref}
\usepackage{authblk}

\title{Dielectric and thermodynamic study of Carbon dioxide (CO$_2$) -  methanol (CH$_3$OH) mixture.  }
\date{}
\author[1]{\small Ra\'ul Fuentes-Azcatl \\
  \href{mailto:razcatl@xanum.uam.mx}{razcatl@xanum.uam.mx}}

\author[2]{\small Hector Dominguez 
  }
\affil[1,2]
{\footnotesize Instituto de Investigaciones en Materiales ,
  Universidad Nacional Aut\'onoma de M\'exico, Circuito Exterior, 04510
  M\'exico D.F., México\\}

\newpage 

\begin{document}

\newpage

\maketitle
\begin{abstract}
In this paper is calculated and analyzed the dielectric behavior of carbon dioxide - methanol mixture and evaluate how the correct reproduction of the electrical properties of carbon dioxide and methanol influences the mixture of these two molecules.\\
 A comparison of two carbon dioxide models and three methanol models is made, where the CO$_{2}$/$\epsilon$ model reproduces and improves the dielectric constant and for the methanol the TraPPE $_{2015}$ and MeOH-4P models improve the dielectric constant $\epsilon$, the surface tension $\gamma$ , and the liquid density $\rho$ of the pure liquid at T=298.15 K and p=1 bar. 
\end{abstract}


\section{Introduction}


The potential for using supercritical fluids for processing a wide range of
solutes has been of increased interest over the past century
\cite{McHugh}. Choosing a supercritical fluid solvent for a particular
task depends on the properties of the solvent at high pressures as well as
the phase behavior exhibited by the solvent-solute mixture at processing 
conditions. Progress in applying high-dielectric, polar supercritical fluids to
processing polar or ionic solutes has not been very successful since, in
general, the critical temperature for a polar solvent such as methanol
(T = 240°C) is high enough to result in very low values for the dielectric
constant. Water is the classic example of a polar solvent that switches to
non-polar behavior at temperatures greater than its critical
temperature (T, = 374°C) \cite{Franck1}. One technique to increase the
polarity or dielectric nature of a supercritical solvent is to add a polar
cosolvent to it.
The objective is to design a mixed solvent which has modest critical
properties as well as reasonable dielectric properties. While a great deal
has been published on the phase behavior of mixtures, only four studies have
been published on the dielectric behavior of binary mixtures at high
pressures.\cite{Diguet}Diguet et al. reported on the dielectric behavior of
methane-methanol mixtures at the very high pressures needed to obtain a
single phase.\cite{Franck2} Deul and Franck investigated the water-benzene
system at both high temperatures and high pressures.\cite{Drake} Drake and
Smith investigated the dielectric behavior of the CO2-argon system and the
CO2-methanol
system at low methanol mole fractions. Dombro et al.\cite{Dombro} reported
on the dielectric behavior of mixtures of carbon dioxide and methanol in the
liquid and mixture-critical regions at 90 to 115°C and pressures to 240 bar.
The CO2-methanol
system is of particular interest from a processing standpoint since these
mixtures have been suggested for use with polar substrates in a wide variety
of potential applications as well as in supercritical fluid chromatography.
However, the effect of methanol on the dielectric behavior of CO2-methanol
mixtures is complicated by the
occurrence of pairwise associates which form between these two
molecules \cite{Hemmaplardh,Fulton }. Modeling and interpreting dielectric
data is a formidable task. For pure liquids, such as methanol, the
Kirkwood-Frohlich equation has been used successfully to interpret dielectric
data \cite{Akhadov,Franck2,Smyth}.\\

Typically the value of the structure correlation parameter for alcohols
is greater than unity since alcohol “monomers’ from linear, hydrogen bonded
multimers, which possess a dipole moment greater than that of a single
isolated alcohol molecule \cite{Prausnitz}. For example, the value of gK
for pure methanol is between 2 and 3, depending upon density and
temperature \cite{Franck2}. Although increasing density promotes the
formation of long-chain multimers, above a certain density
further increases in density decreases G$_k$ since the large linear alcohol
multimers break into shorter chain segments at high densities\cite{Franck2}.
Although the analytical expression suggested by Bottcher et al \cite{Bottcher} has
been used to describe the dielectric behavior of a mixture consisting of
polar and non-polar components, the parameter G$_k$ in that expression is an
empirically fitted function of temperature, density and composition. Also,
this equation does not account for quadrupolar interactions between the
molecules as are found for the
CO$_2$-methanol system.\\
   
One of the challenges of molecular dynamics simulation is the
formulation of realistic potential energy functions describing
molecular interactions in the condensed phase with accurate
force-ﬁeld parameters. The dielectric constant has particular
relevance in solubility processes and for the appropriate
description of the separation of liquid phases in mixtures
having polar components. Unfortunately, nonpolarizable force
ﬁelds are known to have dificulties in reproducing this
property. \cite{Caleman}
New force fields capable of more accurately reproducing electrical properties \cite{spce,tip4pe,nacle,salas,kbre} are helping to better understand mixtures\cite{aljMixture, nacle,kbre,IL}.  \\    

 Regarding CO$_2$, the new force field of Fuentes et al CO$_{2}$/$\epsilon$\cite{rfaCO2} is used in this work, this model reproduces the dielectric constant, surface tension and density at various conditions of pressure and temperature, was also taken into account  the TraPPE model of CO$_2$/TraPPE$_{flex}$\cite{co2Trappe}. For the methanol molecule we used the TraPPE \cite{MetTrappe}  the TraPPE$_{2015}$ \cite{salas} and MeOH-4P\cite{ MeOH-4P }  models, the last two were reparameterized to reproduce the dielectric constant $\epsilon$, the surface tension $\gamma$ , and the liquid density $ \rho$ of the pure liquid at T=298.15 K and p=1 bar. \\

The remaining of the  paper goes as follows. In section 2  
the models are introduced. Section 3
shows the simulation details,Section 4 
 the results are analysed in Section 5. Conclusions
are presented.
\newpage
\section{The Models}
\subsection{Carbon dioxide, CO$_2$}

The two CO$_2$ models used in this work have the same type of potential. The Carbon - Oxygen bond is fixed and the angle formed by the two Oxygens with the Carbon is flexible;  To add flexibility, a harmonic potential is used in the angle, which is described by the three atoms that form the molecule. The flexibility that is added to the potentials has helped to improve the reproduction of various properties at different conditions of pressure and temperature. \cite{fabe, tip4pef}

 \begin{equation}
\label{eqn1}
 U(\theta)\!=\!\frac{k_{\theta}}{2}(\theta-\theta_0)^2 ,
\end{equation}

 \noindent where  $\theta$ is the angle O-C-O and $\theta_0$ referes
 to the equilibrium value,  $k_{\theta}$ is the spring constant.
 Table \ref{table1} shows the potential parameters for the CO$_2$  models used in this work

\begin{table}
\caption{Parameters of the CO$_2$ models considered in this work.}
\label{table1}
\begin{tabular}{|l|ccccccccc|}
\hline\hline
model	&	d $_{OC}$	&	K$_\theta$	&	$\theta_{OCO}$	&	$\epsilon _{O-O} /k_B$	&	$\sigma_{O-O}$	&	$\epsilon _{C-C} /k_B$	&	$\sigma _{C-C}$	&	q$O_O$	&	q$_C$	\\
	&	\AA	&		&	deg	&	K	&	\AA	&	K	&	\AA	&	e	&	e	\\
\hline																			
CO$_2/ \epsilon$	&	1.17	&	500	&	180	&	45.712	&	3.760	&	284.257	&	2.62600	&	-0.3760	&	0.7520	\\

CO$_2$/TraPPE$_{Flex}$	&	1.16	&	1236	&	180	&	79	&	3.05	&	27	&	2.8	&	-0.35	&	0.7	\\

\hline
\end{tabular}
\end{table}

\newpage
\subsection{Methanol, CH$_3$OH}
Table \ref{table2} shows the potential parameters for the methanol models used in this work; all methanol models in this table
use r$_{O-H}$= 0.945 \AA, r$_{O-Me}$= 1.43 \AA,  $\theta$= 108.5$^{\circ}$, ﬂexural constant is k=460.67 kJ mol $^{-1}$ rad $^2$ .

\begin{table}
\caption{Parameters of the CH$_3$OH models considered in this work.}
\label{table2}
\begin{tabular}{|l|ccccccccc|}
\hline\hline
Model	&	        $\epsilon _{O-O} /k_B$  	&	$\sigma_{O-O}$	&	        $\epsilon _{Me-Me} /k_B$  	&	$\sigma_{Me-Me}$	&	q$_O$ /e	&	q$_H$ /e	&	q$_Me$ /e	&	q$_M$ /e	&	l$_{OM}$ / \AA	\\
\hline 
	&		&		&		&		&		&		&		&		&		\\
TraPPE	&	92.9950	&	3.0200	&	97.9980	&	3.7500	&	-0.7000	&	0.4350	&	0.2650	&	-	&	-	\\
TraPPE$_{2015}$	&	83.7142	&	3.0200	&	88.2126	&	3.7500	&	-0.7490	&	0.4650	&	0.2840	&	-	&	-	\\
MeOH-4P	&	90.1180	&	3.1655	&	106.0320	&	3.6370	&	0.0000	&	0.4998	&	0.1546	&	-0.6544	&	-0.037	\\
	&		&		&		&		&		&		&		&		&		\\
\hline
\end{tabular}
\end{table}
 
For the intermolecular potentail between two  molecules the LJ
 and Coulomb interactions are used,

\begin{equation}
\label{ff}
u(r) = 4\epsilon_{\alpha \beta} 
\left[\left(\frac {\sigma_{\alpha \beta}}{r}\right)^{12}-
  \left (\frac{\sigma_{\alpha \beta}}{r}\right)^6\right] +
\frac{1}{4\pi\epsilon_0}\frac{q_{\alpha} q_{\beta}}{r}
\end{equation}

\noindent where $r$ is the distance between sites $\alpha$ and $\beta$,
$q_\alpha$ is the electric charge of site $\alpha$, $\epsilon_0$ is the
permitivity of vacuum,  $\epsilon_{\alpha \beta}$ is the LJ energy scale
and  $\sigma_{\alpha \beta}$ the repulsive diameter for an $\alpha-\beta$ pair.
The cross interactions between unlike atoms are obtained using
the Lorentz-Berthelot mixing rules,

\begin{equation}
\label{lb}
\sigma_{\alpha\beta} = \left(\frac{\sigma_{\alpha\alpha} +
  \sigma_{\beta\beta} }{2}\right);\hspace{1.0cm} \epsilon_{\alpha\beta} =
\left(\epsilon_{\alpha\alpha} \epsilon_{\beta\beta}\right)^{1/2}
\end{equation}

\section{The Simulation Details}

Simulations on binary mixtures, CO$_2$/methanol in the liquid phase were performed using GROMACS~\cite{gromacs} software, version 2018.  In the simulations  864 total molecules were used, 633 CO$_2$ and 231 methanol, which forms a mole fraction of X$_{CH3OH}$=0.267. The NPT ensemble  was used and the equations of motion were solved using the
leapfrog algorithm with a time step of 2 fs with a parameter of 0.5 and  LINCS\cite{Hess2} algorithm to keeps bond distances, using the Nosé-Hoover thermostat and Parrinello-Rahman barostat were applied with parameters of 0.6 and 1.0 ps, respectively  . Electrostatic interactions were handle with the
particle mesh Ewald (PME) method \cite{Essmann} with a grid
space of 0.35 nm and a spline of order 4 and a truncation distance.The cutoff distance was 1.0 nm for both the real part of Coulomb potential
and LJ interactions, within a simulation box of  Lx = Ly = and Lz = 4.39186 nm. The average properties were obtained for at least 30 million configurations (60 ns) after equilibration. The simulation error
for the  dielectric constant  was estimated from the results of three independent simulations.

\section{Results}

\subsection{CO$_2$ - CH$_3$OH mixture}  
  
\subsection*{Density, $\rho$}  

The liquid density is obtained using,
\begin{equation}
\label{dens}
\rho = \frac{M} {<V>}
\end{equation}

where M is the mass of the system and $<V>$ the average volume of the
simulation cell.
For this molecular dynamic study, the mixtures simulations were 
at the concentration of X$_{CH3OH}$=0.267 at isothermal condition of T = 323 K
using different pressures. The choice of the isothermal 
temperature of 323 K and the X $_{CH3OH}$ = 0.267 has been determined by the wealth of experimental data available at this temperature and concentration, including the data used as the input of the simulations.\\

\begin{figure}
\centerline{\psfig{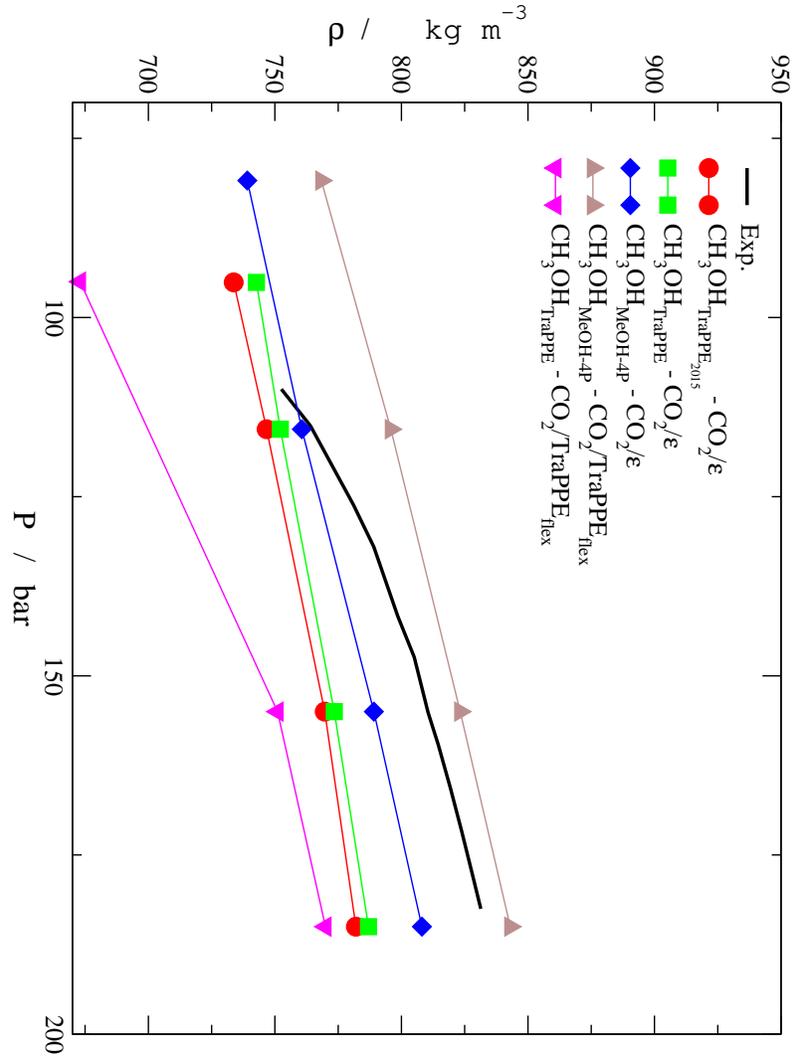}}
\caption{ Density as a function of pressure at 323.15K of temperature. The solid
	black line represents the
	experimental data\cite{B}, red circles the  CH$_3$OH$_{TraPPE_{2015}}$ - CO$_{2}$/$\epsilon$ mixture, green square  the  CH$_3$OH$_{TraPPE}$ - CO$_{2}$/$\epsilon$ mixture, blue diamond the  CH$_3$OH$_{MeOH-4P}$ - CO$_{2}$/$\epsilon$ mixture, brown  triangle up the  CH$_3$OH$_{MeOH-4P}$ - CO$_{2}$/TraPPE$_{Flex}$ mixture and magenta triangle down CH$_3$OH$_{TraPPE}$ - CO$_{2}$/TraPPE$_{Flex}$}
\label{Fig1}
\end{figure}

In figure \ref{Fig1} the total density as function of the pressure is shown
for the different force fields combination. As general trend the
CH$_3$OH$_{MeOH-4P}$ - CO$_{2}$/TraPPE $_{flex}$ has higher values than the other mixtures,
however when the experimental data are compared with the simulations
the  using CO$_{2}$/$\epsilon$  mixtures are slightly better at different pressures.\\

The mixture represented by CH$_3$OH$_{MeOH-4P}$ - CO$_{2}$/$\epsilon$ is more close to the experimental data than the other force fields combination in the region of 115 bars and 130 bars. This result is because the molecule is flexible in the angle, in such a way that it is capable of changing its structure under certain thermodynamic conditions, as will be seen in the structural analysis that follows in this work. 
\subsection*{Dielectric properties}
\begin{figure}
\centerline{\psfig{figure=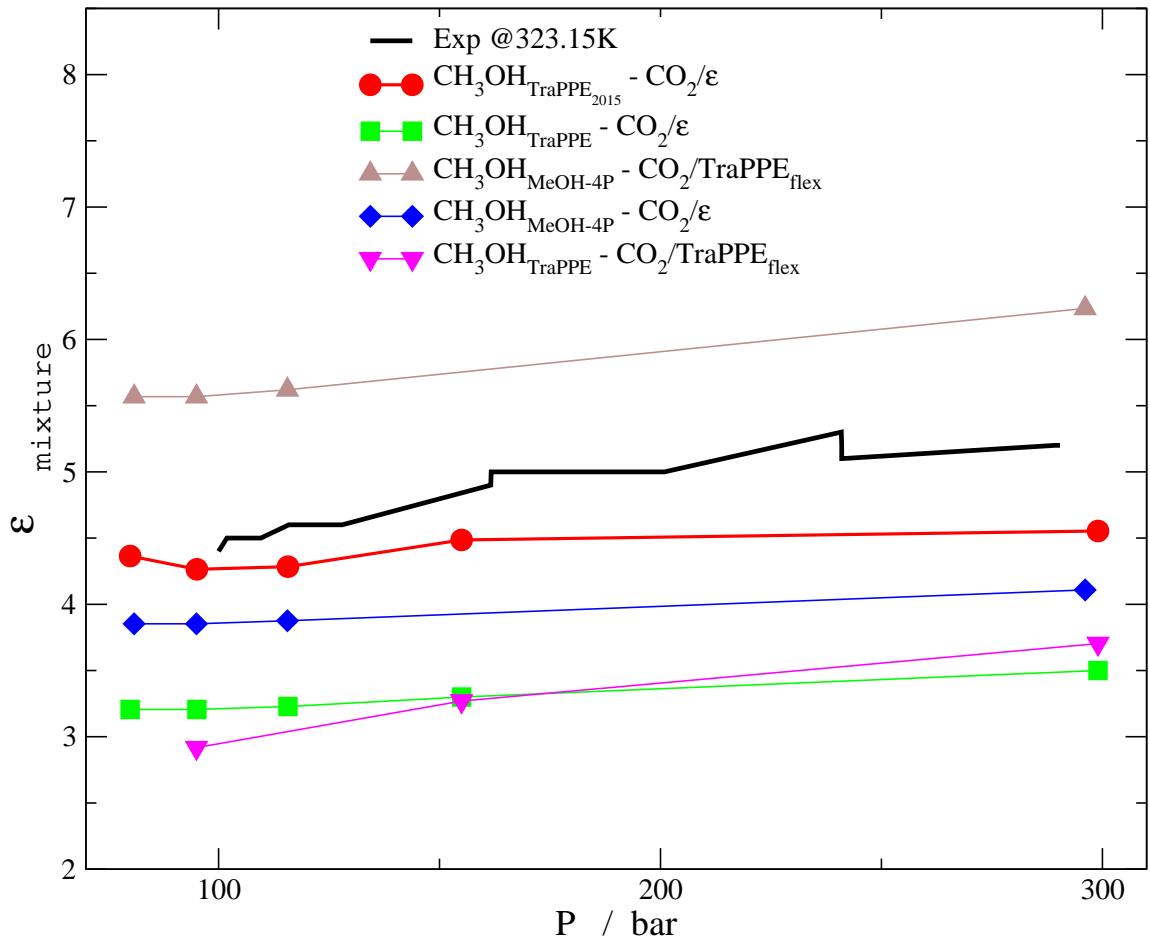,width=14.0cm,angle=-90}}
\caption{ Dielectric constant as a function of pressure at 323.15K of temperature. The solid
	black line represents the
	experimental data\cite{A}, red circles the  CH$_3$OH$_{TraPPE_{2015}}$ - CO$_{2}$/$\epsilon$ mixture, green square  the  CH$_3$OH$_{TraPPE}$ - CO$_{2}$/$\epsilon$ mixture, blue diamond the  CH$_3$OH$_{MeOH-4P}$ - CO$_{2}$/$\epsilon$ mixture, brown  triangle up the  CH$_3$OH$_{MeOH-4P}$ - CO$_{2}$/TraPPE$_{Flex}$ mixture and magenta triangle down CH$_3$OH$_{TraPPE}$ - CO$_{2}$/TraPPE$_{Flex}$ }
\label{Fig2}
\end{figure}

The dielectric constant as function of the pressure
is calculated and plotted in figure \ref{Fig2}. It is
observed in all cases that the dielectric constant increases with the
pressure, however it is noted
that the CH$_3$OH$_{TraPPE_{2015}}$ - CO$_{2}$/$\epsilon$
mixture fits better the experiments than the other models. The CH$_3$OH$_{4sites}$ - CO$_{2}$/TraPPE $_{flex}$ overestimate the value by 10\% and the other combinations underestimate the experimental value by more than 15\%. 
Although the two models CH$_3$OH$_{4sites}$ and CO$_{2}$/$\epsilon$ reproduce the dielectric constant better than others force fields, the combination in this case did not improve the reproduction at these conditions of pressure and temperature, being the combination of the model CH$_3$OH$_{TraPPE_{2015}}$ with CO$_{2}$/$\epsilon$ that is closest to the experimental value. 
The freedom that the CO$_{2}$/$\epsilon$ molecule has in the angle through the harmonic potential, helps to have a better calculation of electrical properties; in recent work\cite{fabe,tip4pef} it has been shown that this type of parameterization helps a closer reproduction of the experimental data. 

The calculations of the dielectric constant was obtained
by the fluctuations~\cite{Neumann} of the total dipole moment {\bf M},

\begin{equation}
\label{Ec4}
\epsilon=1+\frac{4\pi}{3k_BTV} (<{\bf M}^2>-<{\bf M}>^2)
\end{equation}
\noindent where  $k_B$ is the Boltzmann constant,  V and M denote the volume and the total dipole
moment of the simulation box and $T$ the absolute 
temperature. The dielectric constant is obtained for long simulations at constant pressure and temperature, isothermal-isobaric emsamble.

\begin{figure}
\centerline{\psfig{figure=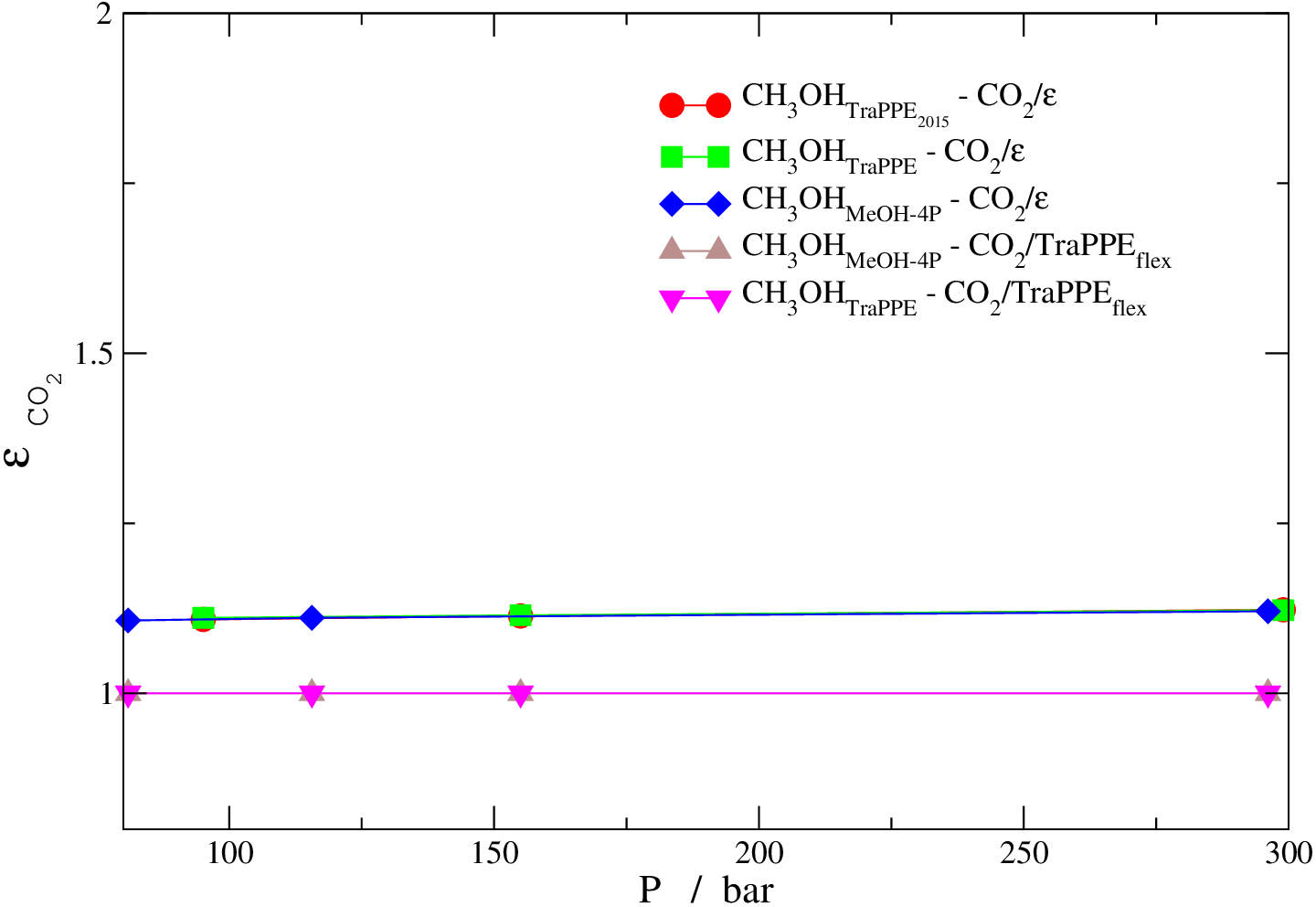,width=12.0cm,angle=0}}
\caption{Dielectric constant of CO$_2$ in the mixture as a function of pressure at 323.15K of temperature. Red circles the  CH$_3$OH$_{TraPPE_{2015}}$ - CO$_{2}$/$\epsilon$ mixture, green square  the  CH$_3$OH$_{TraPPE}$ - CO$_{2}$/$\epsilon$ mixture, blue diamond the  CH$_3$OH$_{MeOH-4P}$ - CO$_{2}$/$\epsilon$ mixture, brown  triangle up the  CH$_3$OH$_{MeOH-4P}$ - CO$_{2}$/TraPPE$_{Flex}$ mixture and magenta triangle down CH$_3$OH$_{TraPPE}$ - CO$_{2}$/TraPPE$_{Flex}$   }
\label{Fig3}
\end{figure}

A relevant part of molecular dynamics is that the molecules can be studied separately in the mixture and thus obtain the average of their dipole moment, as reported in the table \ref{table3}; through that is calculated the dielectric constant of each component, which is induced by the presence of the other component. In the figure \ref{Fig3}, we can see the behavior at different pressures of CO$_2$ component in the mixeture with CH$_3$OH, where independent of the CH$_3$OH model, the flexible CO$_{2}$/$\epsilon$ model describes a dielectric constant closer to the experimental value of 1.3 when it is in an unmixed phase at 323.15 K.

\begin{figure}
\centerline{\psfig{figure=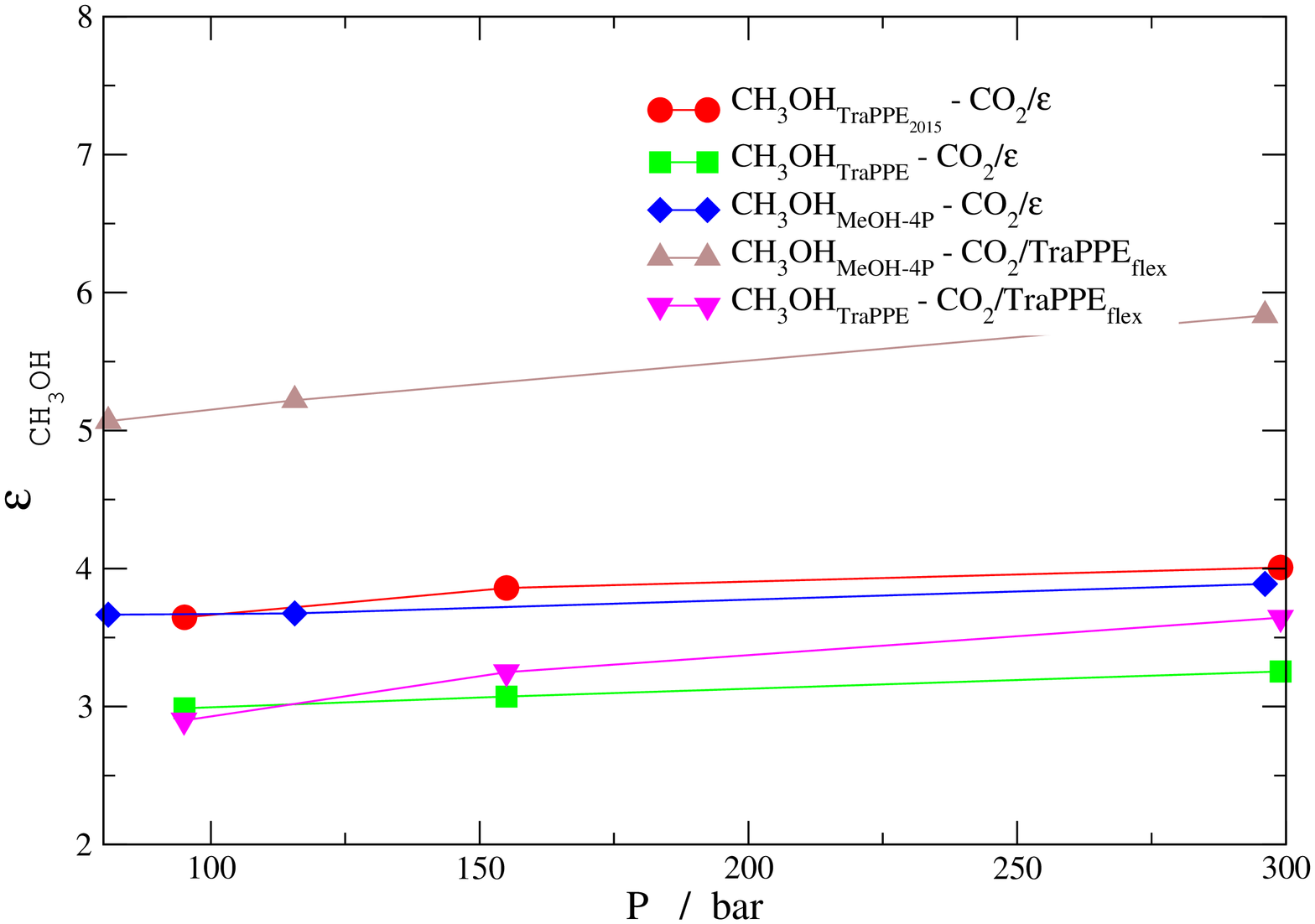,width=12.0cm,angle=-90}}
\caption{Dielectric constant of CH$_3$OH in the mixture as a function of pressure at 323.15K of temperature. Red circles the  CH$_3$OH$_{TraPPE_{2015}}$ - CO$_{2}$/$\epsilon$ mixture, green square  the  CH$_3$OH$_{TraPPE}$ - CO$_{2}$/$\epsilon$ mixture, blue diamond the  CH$_3$OH$_{MeOH-4P}$ - CO$_{2}$/$\epsilon$ mixture, brown  triangle up the  CH$_3$OH$_{MeOH-4P}$ - CO$_{2}$/TraPPE$_{Flex}$ mixture and magenta triangle down CH$_3$OH$_{TraPPE}$ - CO$_{2}$/TraPPE$_{Flex}$    }
\label{Fig4}
\end{figure}

The dielectric constant of the CH$_3$OH in the mixture is presented in the figure \ref{Fig4}, although the model CH$_3$OH$_{4sites}$  is parameterized to reproduce the dielectric constant, it does not have the same dipole moment in the mixture with different models of CO$_2$ as seen in the figure \ref{Fig4} and reported in the table \ref{table3}. When changing the CO$_2$ model, the dielectric constant increases, it is noted that CO$_2$ modifies the average structure and that with the combination of CO$_{2}$/$\epsilon$ and with the models of methanol CH$_3$OH$_{Met-4P}$ and CH$_3$OH$_{TraPPE-2005}$  describe an almost similar dielectric constant, which makes these combinations make the best reproduction of the experimental value, figure \ref{Fig2}

\begin{figure}
\centerline{\psfig{figure=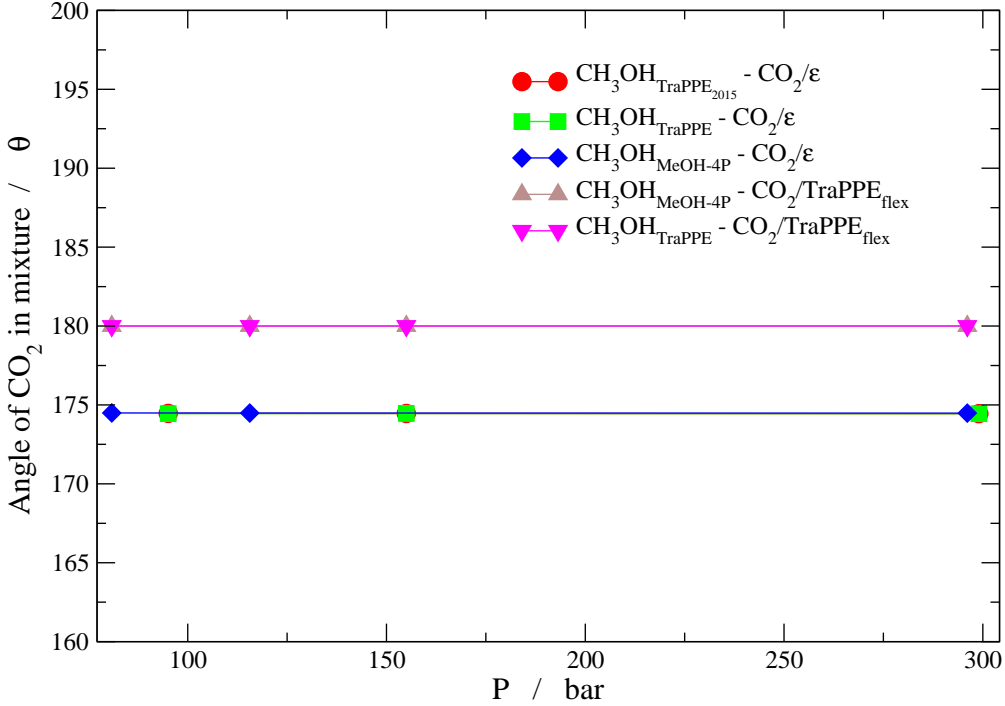,width=12.0cm,angle=-90}}
\caption{ Average angle of CO$_2$ in the mixture as a function of pressure at 323.15K of temperature. Red circles the  CH$_3$OH$_{TraPPE_{2015}}$ - CO$_{2}$/$\epsilon$ mixture, green square  the  CH$_3$OH$_{TraPPE}$ - CO$_{2}$/$\epsilon$ mixture, blue diamond the  CH$_3$OH$_{MeOH-4P}$ - CO$_{2}$/$\epsilon$ mixture, brown  triangle up the  CH$_3$OH$_{MeOH-4P}$ - CO$_{2}$/TraPPE$_{Flex}$ mixture and magenta triangle down CH$_3$OH$_{TraPPE}$ - CO$_{2}$/TraPPE$_{Flex}$   }
\label{Fig5}
\end{figure}

The average angle is plotted  in figure \ref{Fig5}. The value of the
angle is the average value of the distribution in the simulation, and
it is a normal distribution with averages in the value that is
reported. As a general trend, the angle $\theta$ is lower for
the CO$_{2}$/$\epsilon$ model than that for the CO2/TraPPE Flex model that does not change despite containing the harmonic potential in the angle 

The flexibility of the CO$_{2}$/$\epsilon$  give the posibility to the molecule to change the angle, an analysis of this, from 80 to 300 bar of pressure, is represent in the figure \ref{Fig5}. In previous work by Fuentes et al \cite{co2-graf}, was found that under an electric field the CO$_2$ molecule undergoes a change in its structure (angle, dipole moment) and with the results in this work, indicate that the molecule changes its structure what makes the molecule modifies its dipole moment under electrical conditions, either by an electric field or by a polar solvent such as methanol. The change in angle remains constant independent of the occupied methanol model, as seen by using the models  CH$_3$OH$_{MeOH-4P}$, CH$_3$OH$_{TraPPE_{2015}}$ and CH$_3$OH$_{TraPPE}$,then the freedom, in the molecule, to adapt the structure in addition to the reparametrization with target properties helps to better reproduce the experimental values as seen in the tables  \ref{table4} and \ref{table5}.

\newpage

The value in the average of the angle  is related to the
instantaneous values of the polarization of the sample at every step of
simulation.
The dielectric constant with periodic boundary conditions
was calculated from 
\begin{equation}
\label{Ec4}
\epsilon=1+\frac{4\pi}{3k_BTV}N \mu^2G_k
\end{equation}
where G$_k$ is the ﬁnite system Kirkwood factor related to the relative orientation of individual molecular dipoles
\begin{equation}
\label{Ec5}
G_k(r)= <\mu_1 M_1(r) >/ \mu^2
\end{equation}
where $\mu_1$ is the dipole of a reference molecule, 1, at the center
of a sphere of radius r. M$_1$(r) is the sum total of dipoles $\mu_i$ in
the sphere (including the dipole of molecule 1). Local
orientational correlations are averaged out by thermal motion
after the first few coordination shells.The
short-range character of G$_k$(r) was demonstrated by integral
equation theory for the simplest of models for dipolar fluids:
hard spheres with a point dipole fixed at the center.\cite{Stell}  This
relation can be shown to hold for any point dipole system
assuming that the dielectric constant is local. \cite{Fulton}

Is important to specify that the simulations are long at least 60 ns  to get reliable statistics from models having enhanced dipole moments because of the large ﬂuctuations in the dynamics. 
The polarisation factor
G$_K$ (equation \ref{Ec5}) \cite{Glattli}is described in figure \ref{Fig6}, in order to compare if the force
field change its polarisation, is noted that the polarization of the combinations  CH$_3$OH$_{TraPPE_{2015}}$	- CO$_2/\epsilon$ and  CH$_3$OH$_{MeOH-4P}$	- CO$_2/\epsilon$ are similar and this combinations reproduce better the dielectric constant.

\begin{table}
  \caption{Dipole moment for each combination of models of CH$_3$OH and CO$_2$,
    at 323K and 115 bars of temperature and pressure respectively and
    X$_{CH{_4}O}$=0.267}
\label{table3}
\begin{tabular}{|ccccc|}
\hline\hline
 model CH$_3$OH	&	model CO$_2$	&	dm $_{mixture}$	&	dm $_{CH{_3}OH}$ 	&	dm $_{CO{_2}}$	\\
	&		&	debye	&	debye	&	debye	\\
\hline
TraPPE$_{2005}$	&	CO$_2/\epsilon$	&	0.8943	&	2.421	&	0.3371	\\
TraPPE	&	CO$_2/\epsilon$	&	0.8494	&	2.253	&	0.337	\\
MeOH-4P	&	CO$_2/\epsilon$	&	0.8565	&	2.2848	&	0.3348	\\
MeOH-4P	&	CO$_2$/TraPPE$_{flex}$ 	&	0.6109	&	2.2848	&	0	\\
TraPPE		&	CO$_2$/TraPPE$_{flex}$ 	&0.6784	&	2.2512	&	0	\\
\hline
\end{tabular}
\end{table}
\newpage
The argument of the dipole moment can be explained in
terms of the O-C-O angle $\theta$ , the results for the CO$_{2}$/$\epsilon$ force ﬁeld indicate that a dipole moment is induced between the oxygens and the carbon in the
molecule, due to the interaction with methanol independent of the model used.

\begin{figure}
\centerline{\psfig{figure=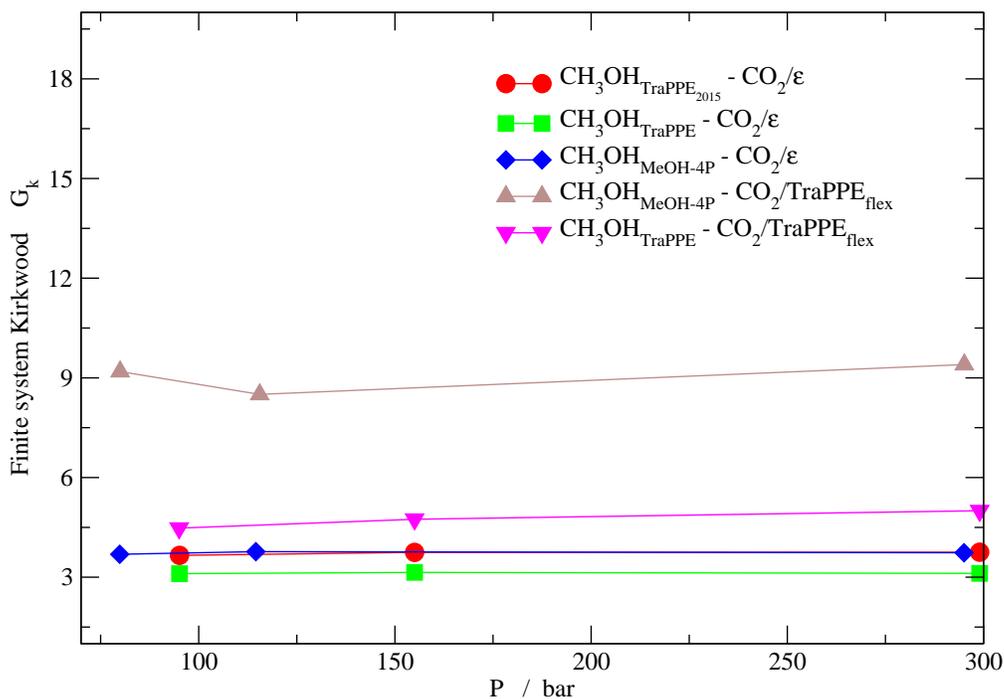,width=12.0cm,angle=-90}}
\caption{ The polarisation factor
	G$_K$ of  the mixture as a function of pressure at 323.15K of temperature. Red circles the  CH$_3$OH$_{TraPPE_{2015}}$ - CO$_{2}$/$\epsilon$ mixture, green square  the  CH$_3$OH$_{TraPPE}$ - CO$_{2}$/$\epsilon$ mixture, blue diamond the  CH$_3$OH$_{MeOH-4P}$ - CO$_{2}$/$\epsilon$ mixture, brown  triangle up the  CH$_3$OH$_{MeOH-4P}$ - CO$_{2}$/TraPPE$_{Flex}$ mixture and magenta triangle down CH$_3$OH$_{TraPPE}$ - CO$_{2}$/TraPPE$_{Flex}$   }
\label{Fig6}
\end{figure}
\newpage
The electric polarization of a polar molecule is made up of orientation, atomic and electric polarization. The orientation polarization can be determined from complex permittivity or so-called dielectric
spectroscopy measurements, with the results in this work help to have more teorical data to compare with the dielectric spectroscopy data.
In figure \ref{Fig6},  the CO$_2$/TraPPE$_{flex}$ molecule induces a greater polarization to the system, that is, to the methanol molecules, which causes the dielectric constant to be underestimated by around 20$\%$. In the case of the combination of CO$_{2}$/$\epsilon$ and CO$_3$OH/TraPPE there is a slight decrease in the polarization of the system, but because the CH$_3$OH/TraPPE does not reproduce the dielectric constant \cite {salas} the calculated value is underestimated by almost 80$\%$ figure \ref{Fig2}.

\subsection*{Self Diffusion coefficient}

The diffusion coefficient was obtained from the
long-time limit of the mean square displacement according
to the Einstein relation \cite{Allen},

\begin{equation}
\label{EcD}
D= \lim _{x\to\infty} <(r(t)-r(0))^2> / 6t
\end{equation}
where r(t) corresponds to the position vector of the center of
mass at time t and the averaging $<...>$  is performed over
both time origins.

\begin{figure}
\centerline{\psfig{figure=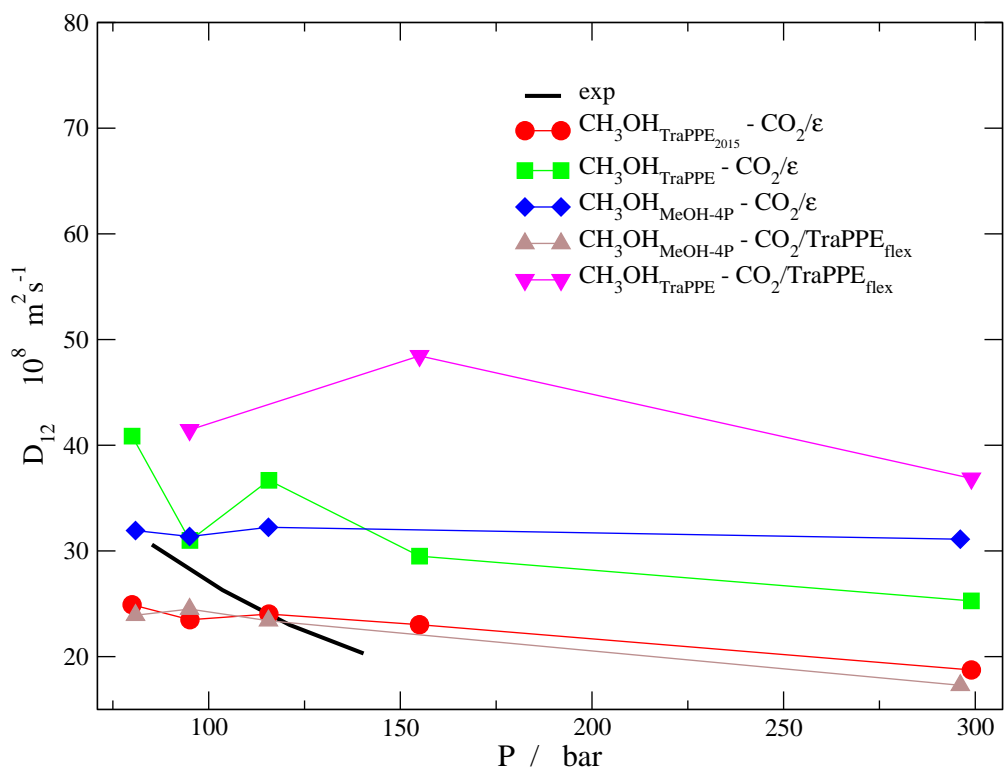,width=12.0cm,angle=-90}}
\caption{ The self Diffusion coefficient D  in the mixture as a function of pressure at 323.15K of temperature. The solid
	black line represents the
	experimental data \cite{1}, red circles the  CH$_3$OH$_{TraPPE_{2015}}$ - CO$_{2}$/$\epsilon$ mixture, green square  the  CH$_3$OH$_{TraPPE}$ - CO$_{2}$/$\epsilon$ mixture, blue diamond the  CH$_3$OH$_{MeOH-4P}$ - CO$_{2}$/$\epsilon$ mixture, brown  triangle up the  CH$_3$OH$_{MeOH-4P}$ - CO$_{2}$/TraPPE$_{Flex}$ mixture and magenta triangle down CH$_3$OH$_{TraPPE}$ - CO$_{2}$/TraPPE$_{Flex}$  }
\label{Fig6}
\end{figure}

The self Diffusion coefficients are also study in the mixtures, figure \ref{Fig6} show the different features, the CH$_3$OH$_{TraPPE_{2015}}$/CO$_2/\epsilon$  and
CH$_3$OH$_{MeOH-4P}$/CO$_2$/$TraPPe$ mixtures have similar values and they reproduce the experimental data\cite{1} at 115 bar but at lower pressures the values have a difference of more than 10 percent.
The  CH$_3$OH$_{MeOH-4P}$/CO$_2/\epsilon$ have a percent of difference in the 95.1 bar but at high pressures deviates from the reproduction of the experimental values. The other combination are further away from the experimental values,

In tables \ref{table4} and \ref{table5} are  compared the combination of the CO$_2$ and CH$_3$OH in the mixture with X$_CH_{3}OH$=0.257 respect to the experimental data that can reproduce follow the procedure used to qualify water models \cite{tip4pef} for a certain property the prediction of the mixture model adopts the value X, and the experimental value is X$_{exp}$. The score is calculated by the equation \ref{qual}
\begin{equation}
\label{qual}
M= min\left\{anint\left[10 -  abs\left(\frac{X-X_{exp}}{X_{exp}   \textit{tol}}\right)\right], 0 \right\},
\end{equation}
where the tolerance \textit{tol} is given as a percentage and anint is the nearest integer function.

The values that reproduce the force fields CH$_3$OH$_{TraPPE_{2015}}$ - CO$_2/\epsilon$ are the closest to the experimental value, taking into account a tolerance of 2 percent for the density and dielectric constant and a 5 percent tolerance for diffusion, which takes into account the error that this property has in the simulation.

\begin{table}
	\caption{Experimental and simulation data of different CO$_{2}$-CH $_{3}$OH force fields in the mixture. Thermodynamic conditions as reported in each entry.}
	\label{table4}
	\scalebox{0.6}[0.8]{
	\begin{tabular}{|lcccccccc|c|}
		\hline\hline
	&		&		&		&		&		&		&		&		\\
Force fields	&		&	CO$_{2}$/$\epsilon$ 	&	CO$_{2}$/$\epsilon$	&	CO$_{2}$/$\epsilon$	&		&	CO$_{2}$/$\epsilon$ , 	&	CO$_{2}$/$\epsilon$	&	CO$_{2}$/$\epsilon$	\\
&		&	{\small CH $_{3}$OH $_{TraPPE _{2015}}$}	&	{\small CH $_{3}$OH $_{TraPPE}$}	&	{\small CH $_{3}$OH $_{MeOH-4P}$}	&		&	{\small CH $_{3}$OH $_{TraPPE _{2015}}$}	&	{\small CH $_{3}$OH $_{TraPPE}$}	&	{\small CH $_{3}$OH $_{MeOH-4P}$}	\\
\hline																	
Property	&	Exp.	&	Quantity	&	Quantity	&	Quantity	&	Tol. (\%)	&	 Score 	&	 Score 	&	 Score 	\\
\hline																	
\multicolumn{9}{|l|}{Densities at 323.15K }\\																	
$\rho/g cm^{-3}$ [115.6 bar]	&	765	&	746.7	&	752.152	&	760.634	&	2	&	8.80	&	9.16	&	9.71	\\
$\rho/g cm^{-3}$ [155 bar]	&	810.36	&	769.712	&	773.485	&	789.12	&	2	&	7.49	&	7.72	&	8.69	\\
\hline																	
\multicolumn{9}{|l|}{Static dielectric constant   at 323.15K}\\																	
$\varepsilon$[100 bar]	&	4.41	&	4.27	&	3.22	&	5.59	&	2	&	8.41	&	0.00	&	0.00	\\
$\varepsilon$[115.6 bar]	&	4.57	&	4.27	&	3.22	&	5.63	&	2	&	6.72	&	0.00	&	0.00	\\
$\varepsilon$[155 bar]	&	4.82	&	4.49	&	3.33	&	5.77	&	2	&	6.58	&	0.00	&	0.15	\\
$\varepsilon$[280 bar]	&	5.19	&	4.57	&	3.46	&	6.19	&	2	&	4.03	&	0.00	&	0.37	\\
\hline																	
\multicolumn{9}{|l|}{Self-diffusion coefficient/cm$^2$s$^{-1}$ at 323.15K}\\																	
																	
D$_[95 bar]$	&	28.2	&	23.43	&	30.64	&	31.37	&	5	&	6.62	&	8.27	&	7.75	\\
D$_[115 bar]$	&	24.05	&	24.04	&	36.7	&	32	&	5	&	9.99	&	0.00	&	3.39	\\
\hline																	
\multicolumn{6}{|l|}{Overall score (out of 10)}								&	7.33	&	3.14 &	3.76		\\
\hline

\end{tabular}}
\end{table}

\begin{table}
	\caption{Experimental and simulation data of different CO$_{2}$-CH $_{3}$OH force fields in the mixture. Thermodynamic conditions as reported in each entry.}
	\label{table5}
	\scalebox{0.6}[0.8]{
		\begin{tabular}{|lcccccc|c|}
			\hline\hline

	&		&		&		&		&		&		\\
Force fields	&		&	{\small CO$_{2}$/TraPPE$ _{Flex}$ }	&	{\small CO$_{2}$/TraPPE$ _{Flex}$}	&		&	{\small CO$_{2}$/TraPPE$ _{Flex}$ }	&	{\small CO$_{2}$/TraPPE$ _{Flex}$}	\\
&		&	{\small CH $_{3}$OH $_{MeOH-4P}$}	&	{\small CH $_{3}$OH $_{TraPPE}$}	&		&	{\small CH $_{3}$OH $_{MeOH-4P}$}	&	{\small CH $_{3}$OH $_{TraPPE}$}	\\
\hline													
Property	&	Exp.	&	Quantity	&	Quantity	&	Tol. (\%)	&	 Score 	&	 Score 	\\
\hline													
\multicolumn{7}{|l|}{Densities at 323.15K }\\													
$\rho/g cm^{-3}$ [115 bar]	&	765	&	796.44	&	700.3	&	2	&	7.95	&	5.77	\\
$\rho/g cm^{-3}$ [155 bar]	&	810.36	&	823.38	&	751.9	&	2	&	9.20	&	6.39	\\
\hline													
\multicolumn{7}{|l|}{Static dielectric constant   at 323.15K}\\													
$\varepsilon$[100 bar]	&	4.41	&	5.58	&	2.96	&	2	&	0.00	&	0.00	\\
$\varepsilon$[115.6 bar]	&	4.57	&	5.63	&	3.05	&	2	&	0.00	&	0.00	\\
$\varepsilon$[155 bar]	&	4.82	&	5.75	&	3.26	&	2	&	0.35	&	0.00	\\
$\varepsilon$[280 bar]	&	5.19	&	6.18	&	3.65	&	2	&	0.46	&	0.00	\\
\hline													
\multicolumn{7}{|l|}{Self-diffusion coefficient/cm$^2$s$^{-1}$ at 323.15K}\\													
													
D$_[95 bar]$	&	28.2	&	24.47	&	41.41	&	5	&	7.35	&	0.63	\\
D$_[115 bar]$	&	24.05	&	23.42	&	43.92	&	5	&	9.48	&	0.00	\\
\hline													
\multicolumn{5}{|l|}{Overall score (out of 10)}									&	4.35	&	1.60	\\
\hline

\end{tabular}}
\end{table}

\newpage

\section{Conclusions}
This paper present the calculations of the dielectric behavior
CO2-methanol mixtures at different termodynamics conditions.   In previous work by Fuentes et al \cite{co2-graf}, was found that under an electric field the CO2 molecule undergoes a change in its structure  and therefore the dipole moment. The results in this paper indicate that the molecule changes its structure what makes the molecule modifies its dipole moment due to contact with a polar solvent such as methanol as happened under electrical conditions.
The force fields CH$_3$OH$_{TraPPE_{2015}}$ - CO$_2/\epsilon$ transfer the values they reproduce to the mix and this helps to have results closer to the experimental data. These force fields have been parameterized to reproduce the dielectric constant, surface tension and the density.\\
The above indicates that the formulation of realistic potential energy functions describing molecular interactions in the condensed phase with accurate force-ﬁeld parameters allows the study of mixtures with a better approximation to the experimental data. The dielectric constant has particular relevance in solubility processes of liquid phases in mixtures having polar components.

\section {Acknowledgements}
 
RFA thanks DGAPA-UNAM for a postdoctoral fellowship. and also thank the SECTEI of Mexico city for financial support.

\end{document}